\documentclass[preprint,nofootinbib,amsmath,amssymb,aps,preprintnumbers,superscriptaddress]{revtex4-1}
\usepackage[dvipdfmx]{graphicx}
\usepackage{bm}
\usepackage{braket}
\usepackage{cases} 
\usepackage{here}
\usepackage{color}
\usepackage{upgreek}
\usepackage{ulem}
\usepackage{empheq}

\usepackage{dcolumn}

\begin{document}

\preprint{NCTS-TH/2013}

\title{Inertial and gravitational effects on a geonium atom}

\author{Asuka Ito}
\email[]{asuka.i.aa@m.titech.ac.jp}
\affiliation{Physics Division, National Center for Theoretical Sciences, Hsinchu, 30013, Taiwan}


\begin{abstract}
We reveal all linear order
inertial and gravitational effects on a 
non-relativistic Dirac particle (mass $m$) on the Earth
up to the order of $1/m$ in the Foldy-Wouthuysen-like expansion. 
Applying the result to 
Penning trap experiments where a Dirac particle experiences 
the cyclotron motion and the spin precession
in a cavity, i.e., a geonium atom, we study
modifications to the $g$-factor of such as the electron.
It is shown that each correction from gravity has 
different dependence on the cyclotron frequency and the mass $m$.
Therefore, their magnitude change depending on situations.
In a particular case of an electron $g$-factor measurement, 
the dominant correction to the observed $g$-factor
comes from effects of the Earth's rotation,
which is $\delta g / 2 \simeq 5.2 \times 10^{-17}$.
It may be detectable in the near future.
\end{abstract}

\maketitle

\tableofcontents
\newpage
\section{Introduction}
In order to confirm predictions from the standard model of particle 
physics~\cite{Aoyama:2019ryr,Aoyama:2020ynm}
and/or probe beyond the standard 
model~\cite{Giudice:2012ms,Czarnecki:2001pv}, magnetic moments/$g$-factors of fermions
have been measured intensively.
For instance, measurements for
the electron $g$-factor~\cite{Hanneke:2008tm,Odom:2006zz,VanDyck:1987ay}
and
the muon $g$-factor~\cite{Muong-2:2021ojo,Bennett:2006fi,Bailey:1978mn} 
have been conducted with very high accuracy.
For the case of the electron, 
one can resolve a one-electron quantum transition
in current quantum optical 
technologies~\cite{Brown:1985rh,DUrso:2003ilf,Hanneke:2010au} and
it enables us to measure the electron $g$-factor with remarkable small
uncertainty, which is $2.8 \times 10^{-13}$ for $g/2$.
Considering this current sensitivity,
effects of gravity on the $g$-factor measurements
may not be negligible.
Furthermore, discrepancies between theoretical predictions and experimental results were reported 
both for the electron~\cite{Hanneke:2008tm} and the muon~\cite{Muong-2:2021ojo}.
Therefore it would be important to explore the possibility that effects of gravity could reconcile the
discrepancies.

Actually, gravitational effects on the \textit{g}-factor of the 
electron or the muon have been studied 
intensively~\cite{Morishima:2018bqz,Visser:2018omi,Nikolic:2018dap,Venhoek:2018biz,Laszlo:2018llb,Jentschura:2018mlv,Ulbricht:2019dzm}.
In order to investigate effects of gravity, 
it is crucial to consider the equivalence principle appropriately, namely
we need to use a coordinate moving with an observer bound on the 
surface of the Earth~\cite{Ni:1978zz,Ito:2020wxi}.
This aspect was emphasized 
qualitatively in~\cite{Visser:2018omi,Nikolic:2018dap,Venhoek:2018biz}
and investigated quantitatively 
in~\cite{Laszlo:2018llb,Notari:2019qcx,Ulbricht:2019dzm}.
\cite{Laszlo:2018llb} and \cite{Notari:2019qcx} analyzed some
general relativistic effects
with the use of the Fermi-Walker transport in equations of motion
for the case of the muon.
\cite{Ulbricht:2019dzm} evaluated an inertial effect,
an acceleration relative to (local) inertial frames, which is 
characterized by $|\bm{a}| = 9.81 \, {\rm m/s}^{2}$
in the case of the gravity of Earth, in 
a Hamiltonian of a Dirac particle by taking the non-relativistic limit 
for the case of the electron.
However in \cite{Ulbricht:2019dzm}, other effects of Earth's gravity were missed.
Moreover, \cite{Ulbricht:2019dzm} has not studied a spin-orbit coupling induced by $\bm{a}$, which 
actually gives rise to a leading correction among corrections from $\bm{a}$ as we will see.
In this paper, 
we study all linear order general relativistic effects, namely
inertial effects of the acceleration, the rotation due to the Earth
and tidal effects due to weak gravitational fields.
In particular, we evaluate magnitude of the general relativistic corrections
for the case of the electron \textit{g}-factor measurements.

To this end, we first take the non-relativistic limit of a Hamiltonian 
for a Dirac particle (mass $m$) up to the order of $1/m$
in the Foldy-Wouthuysen-like expansion~\cite{Foldy:1949wa,Bjorken:1965zz,Ito:2020wxi} 
on a proper reference frame~\cite{Ni:1978zz,Ito:2020wxi}.
We mention that analyzing the Hamiltonian is more useful than equations of motion 
because it enables us to access special effects like a spin-orbit coupling, which 
can not be derived in equations of motion.
Next, we apply the obtained Hamiltonian to the situation of
Penning trap experiments where a Dirac particle experiences 
the cyclotron motion the spin precession
in a cavity, i.e., a geonium atom, and estimate
magnitude of the effects of gravity.
As a result, it turns out that
effects of the Earth's rotation is dominant among
the general relativistic corrections.
It can be detected if 
the current sensitivity is improved by 4 orders of magnitude.

The paper is organized as follows.
In the section \ref{proref},
we introduce a proper reference frame and 
consider the Dirac equation in the coordinate.
Then a Hamiltonian in the proper reference frame is obtained.
In the section \ref{nonrela}, we take the non-relativistic limit
of the Hamiltonian up to the order of $1/m$.
This manifests all linear order 
inertial and gravitational effects on a non-relativistic Dirac particle.
In the section \ref{eacheffect},
we apply the non-relativistic Hamiltonian to the case of Penning trap
experiments and analyze the inertial and gravitational effects on
the cyclotron motion and the spin precession.
In the section \ref{detectability},
we consider a particular case of an electron $g$-factor measurement
to probe the detectability of the general relativistic corrections.
The final section is devoted to the conclusion.
In the appendix \ref{review},
a brief review of Penning trap experiments is given
for reference.
\section{Dirac equation in a proper reference frame}
\label{proref}
In this section, we investigate inertial and gravitational effects on 
a Dirac particle bound on the surface of the Earth perturbatively.
To this end, we use a proper reference frame~\cite{Ni:1978zz,Ito:2020wxi}.
A proper reference frame for an observer who is accelerating 
against the center of the Earth, 
$|\bm{a}| =  9.81 \, {\rm m/s}^{2}$, and rotating
due to the Earth's rotation, 
$|\bm{\omega}| = 7.27 \times 10^{-5} \, {\rm rad/s}$,
relative to (local) inertial frames can be 
constructed with the use of the Fermi-Walker 
transport~\cite{Ni:1978zz,Ito:2020wxi}.
Then the metric in the frame is obtained perturbatively in powers of 
$ax \ll 1$, $\omega x \ll 1$ and
(the Riemann tensor $\times xx$) $\ll$ 1, where 
$x$ represents a typical scale of a system.
In this paper, we use the term ``inertial'' for $a_{i}$ and $\omega_{i}$, 
and ``gravitational'' for the curvature. 
%
Up to the quadratic order for $x$, the metric is given by
%
%
\begin{empheq}[left=\empheqlbrace]{align} 
  g_{00} &= - 1 -2 a_{i} x^{i} - R_{0i0j} x^{i} x^{j} , \nonumber \\
  \label{met033}
  g_{0i} &= - \omega_{k} \epsilon_{0ijk} x^{j} -\frac{2}{3} R_{0jik} x^{j} x^{k} ,   \\        
  g_{ij} &= \delta_{ij} - \frac{1}{3} R_{ikjl} x^{k} x^{l} ,   \nonumber
\end{empheq}
where the anti symmetric tensor is assigned as $\epsilon_{0123} = 1$.
The Riemann tensor is evaluated at $\bm{x} = 0$, so that
it only depends on time $x^{0}$. 
At this occasion, 
we have not specified the source of the curvature.
The origin of the spatial coordinates
is set on the center of gravity
of a system, which traces a worldline of a freely falling particle in the limit of $\bm{a} = \bm{\omega} = \bm{0}$.
In the case of a geonium atom, 
the origin should be
at the center of the cyclotron motion explained in the 
appendix \ref{review}.
%

We now consider the Dirac equation in the proper reference frame by 
using the metric (\ref{met033}).
The Dirac equation in curved spacetime is given by~\cite{Birrell:1982ix}
\begin{equation}
  i \gamma^{\hat{\alpha}} e^{\mu}_{\hat{\alpha}} \left( \partial_{\mu} - \Gamma_{\mu} 
                - i e A_{\mu} \right) \psi = m \psi \ ,  \label{dira}
\end{equation}
where $\gamma^{\hat{\alpha}}$,  $e$,  $m$, $A_{\mu}$ are the gamma matrices, 
an electromagnetic charge, a mass
and a vector potential, respectively.  
The tetrad $e^{\mu}_{\hat{\alpha}}$ is defined to satisfy
\begin{equation}
e^{\hat{\alpha}}_{\mu} e^{\hat{\beta}}_{\nu} \eta_{\hat{\alpha}\hat{\beta}} = g_{\mu\nu} \ . \label{tetrad}
\end{equation}
Note that  $\eta_{\hat{\alpha}\hat{\beta}}$ is the Minkowski metric of a local inertial frame and
hat is used for the frame.
More explicitly, for the metric (\ref{met033}), 
the tetrads are constructed as
\begin{equation}
  e^{\hat{\alpha}}_{0} = \delta^{\hat{\alpha}}_{0} \left( 1 + a_{i} x^{i} \right) 
           - \frac{1}{2} \delta^{\hat{\alpha}}_{\alpha}  R^{\alpha}_{\ k0l}\ , 
  \quad
  e^{\hat{\alpha}}_{i} =  \delta^{\hat{\alpha}}_{0} 
                           \omega_{k} \epsilon_{0ijk} x^{j} 
                         + \delta^{\hat{\alpha}}_{i}
        - \frac{1}{6} \delta^{\hat{\alpha}}_{\alpha}  R^{\alpha}_{\ kil}x^{k} x^{l}c  \ .
\label{tet}
\end{equation}
The spin connection is defined by
\begin{equation}
  \Gamma_{\mu} 
  = -\frac{i}{2} e^{\hat{\alpha}}_{\nu} \sigma_{\hat{\alpha}\hat{\beta}} 
        \left( \partial_{\mu} e^{\nu\hat{\beta}} + \Gamma^{\nu}_{\lambda\mu} e^{\lambda\hat{\beta}} \right), 
        \label{spicone}
\end{equation}
where
$\sigma_{\hat{\alpha}\hat{\beta}} = \frac{i}{4} [ \gamma_{\hat{\alpha}}, \gamma_{\hat{\beta}} ] $ is 
a generator of the Lorentz group and 
$\Gamma^{\mu}_{\nu\lambda}$ is the Christoffel symbol.
For the metric (\ref{met033}), the spin connection
at the linear order for inertial and gravitational terms 
can be calculated as follows:
\begin{empheq}[left=\empheqlbrace]{align}
 \label{spi0} 
 \Gamma_{0} &= -\frac{1}{2} \gamma^{\hat{0}} \gamma^{\hat{i}} a_{i} 
            - \frac{1}{4} \gamma^{\hat{i}} \gamma^{\hat{j}} \omega_{k} \epsilon_{0ijk}
       - \frac{1}{2} \gamma^{\hat{0}} \gamma^{\hat{i}} R_{0i0j} x^{j}
                - \frac{1}{4} \gamma^{\hat{i}} \gamma^{\hat{j}} R_{ij0k} x^{k}
        \ ,  \quad  \\
 \label{spii}  
 \Gamma_{i} &= 
              - \frac{1}{2} \gamma^{\hat{0}} \gamma^{\hat{j}}
               \omega_{k} \epsilon_{0ijk}
              -
             \frac{1}{4} \gamma^{\hat{0}} \gamma^{\hat{j}} R_{0jik} x^{k} 
              - \frac{1}{8} \gamma^{\hat{j}} \gamma^{\hat{k}} R_{jkil} x^{l} 
               \ .         
\end{empheq}
Here we have rewritten $\delta_{\hat{\alpha}}^{\mu} \gamma^{\hat{\alpha}}$ as $\gamma^{\hat{\mu}}$ and 
we will do so throughout.

On the other hand,
the Dirac equation (\ref{dira}) can be rewritten as
\begin{eqnarray}
 i \gamma^{0} \partial_{0} \psi &=& \left[ i \gamma^{0} \left( \Gamma_{0} + i e A_{0} \right)
                                  - i \gamma^{j} \left( \partial_{j} - \Gamma_{j} - i e A_{j} \right)  
                                  +  m   \right]  \psi   \nonumber \\
                 &=&   \gamma^{0} H \psi \ ,
\end{eqnarray}
where we defined a Hamiltonian $H$ and the gamma matrices in curved spacetime,
$\gamma^{\mu} = e^{\mu}_{\hat{\alpha}} \gamma^{\hat{\alpha}}$,  
satisfying the relation  
\begin{equation}
  \{ \gamma^{\mu} ,  \gamma^{\nu} \}  =  - 2 g^{\mu\nu} \ .
\end{equation}
Let us express the Hamiltonian in terms of the gamma matrices of the local inertial frame
instead of those of curved spacetime.
Because of $\gamma^{0}\gamma^{0} = -g^{00}$, we obtain
\begin{equation}
  H =  (g^{00})^{-1}  \left[ i g^{00} \left( \Gamma_{0} + i e A_{0} \right)
                                  + i \gamma^{0}\gamma^{j} \left( \partial_{j} - \Gamma_{j} - i e A_{j} \right)  
                                  -  \gamma^{0} m   \right]  \ .   \label{hamihami}
\end{equation}
%
%
%
%
%
%
%
%
%
%
Using Eqs.\,(\ref{met033}) and (\ref{tet}), we calculate
\begin{eqnarray}
  (g^{00})^{-1}  \gamma^{0} \gamma^{j}  
       &\simeq&
        - \gamma^{\hat{0}} \gamma^{\hat{j}} 
                    \left( 1 + a_{i} x^{i} \right)
        + \gamma^{\hat{i}} \gamma^{\hat{j}} \omega_{k} 
               \epsilon_{0ilk} x^{l}
            -  \gamma^{\hat{0}} \gamma^{\hat{j}} 
            - \frac{1}{2} R_{0kjl} x^{k} x^{l} 
             \nonumber \\
         && - \frac{1}{6} \gamma^{\hat{0}} \gamma^{\hat{a}} R_{jkal}  x^{k} x^{l} 
           - \frac{1}{2} \gamma^{\hat{0}} \gamma^{\hat{j}} R_{0k0l} x^{k}x^{l} 
            + \frac{1}{6} \gamma^{\hat{a}} \gamma^{\hat{j}} R_{ak0l} x^{k} x^{l}  \ .  \label{000j}
\end{eqnarray}
Similarly, we have
\begin{equation}
  (g^{00})^{-1} \gamma^{0} \simeq 
    - \gamma^{\hat{0}} \left( 1 + a_{i} x^{i} \right) 
    - \gamma^{\hat{i}}  \omega_{k}  \epsilon_{0ilk} x^{l}
    - \frac{1}{2} \gamma^{\hat{0}} R_{0k0l} x^{k}x^{l}
    + \frac{1}{6} \gamma^{\hat{a}} R_{ak0l} x^{k} x^{l} \ .   \label{000}
\end{equation}
Therefore using Eqs.\,(\ref{spi0}), (\ref{spii}),
(\ref{000j}) and (\ref{000}) in the Hamiltonian
(\ref{hamihami}), we obtain
\begin{eqnarray}
 H &=&  - \frac{i}{2} \gamma^{\hat{0}} \gamma^{\hat{i}}  \left( a_{i}
         + R_{0i0j} x^{j} \right)
         - \frac{i}{4} \gamma^{\hat{i}} \gamma^{\hat{j}} R_{0ikj} x^{k}
    -  \frac{i}{8}  \gamma^{\hat{0}} \gamma^{\hat{i}} \gamma^{\hat{j}} \gamma^{\hat{k}} R_{jkil} x^{l}
            - eA_{0}        \nonumber \\
    && + \Big[  \gamma^{\hat{0}} \gamma^{\hat{i}} 
         \Big( \delta^{j}_{i} \left( 1 + a_{i}x^{i}  \right) + \theta^{j}_{i} \Big)
         - \gamma^{\hat{i}} \gamma^{\hat{j}} \left(
         \omega_{k} \epsilon_{0ilk} x^{l}  
         + \frac{1}{6}  R_{ik0l} x^{k} x^{l}  \right)
         + \frac{1}{2} R_{0kjl} x^{k} x^{l}
          \Big]  
         \left( -i \partial_{j} -  eA_{j} \right) \nonumber \\
    &&  +  \left[ \gamma^{\hat{0}} 
           \left(1 + a_{i}x^{i} + \frac{1}{2} R_{0k0l} x^{k} x^{l} \right)
        -  \gamma^{\hat{i}} \left( \omega_{k} \epsilon_{0ijk} x^{j} 
        + \frac{1}{6}   R_{ik0l} x^{k} x^{l} \right)
    \right]  m   \ ,  \label{HHH}
\end{eqnarray}
where we defined
\begin{equation}
  \theta^{j}_{i} = 
                \frac{1}{2} \delta^{j}_{i} R_{0k0l} x^{k}x^{l}
              + \frac{1}{6}  R_{jkil}  x^{k} x^{l}  \ .
\end{equation}
The above Hamiltonian is a 4$\times$4 matrix and contains both the fermion and the anti-fermion.
What we want to consider is the fermion particle with a non-relativistic velocity.
In order to take the non-relativistic limit of the Hamiltonian for the fermion, 
we need to separate the fermion and the anti-fermion while expanding the Hamiltonian
in powers of $1/m$.
In the next section, we will explicitly show how to perform this.
\section{Non-relativistic limit of the Hamiltonian}
\label{nonrela}
In the previous section, the (non-relativistic) Hamiltonian of 
a Dirac field in the proper reference frame was derived.
We take the non-relativistic limit of the Hamiltonian (\ref{HHH})
on the assumption that a fermion has a velocity well below the speed of light,
which is usual in experiments on the Earth like
the electron \textit{g}-factor measurements~\cite{Hanneke:2008tm,Odom:2006zz}.

The Hamiltonian (\ref{HHH}) can be divided into the even part, the odd part and 
the terms multiplied by $m$:
\begin{eqnarray}
 H &=&   -\frac{i}{2} \alpha^{i} \left( a_{i}
         + R_{0i0j} x^{j} \right) 
         + \frac{i}{8}  \alpha^{i} \alpha^{j} \alpha^{k}  R_{jkil} x^{l}
         + \alpha^{j} 
         \Big( \delta^{j}_{i} \left( 1 + a_{i}x^{i}  \right) + \theta^{j}_{i} \Big)
          \Pi_{j}
\nonumber \\
    &&   - eA_{0}
         - \frac{i}{4} \alpha^{i} \alpha^{j} 
           \left(  \omega_{k} \epsilon_{0ijk} - R_{0ikj} x^{k} \right)
         +\bigg[  
          \frac{1}{2} R_{0kjl} x^{k} x^{l}
         + \alpha^{i} \alpha^{j} \left( \omega_{k} \epsilon_{0ilk} x^{l} 
         + \frac{1}{6} R_{ik0l} x^{k} x^{l} \right)  \bigg]  
         \Pi_{j}
 \nonumber \\
    &&  +  \left[ \beta 
           \left(1 + a_{i}x^{i} + \frac{1}{2} R_{0k0l} x^{k} x^{l} \right)
        -  \beta \alpha^{i} \left( \omega_{k} \epsilon_{0ijk} x^{j} 
           + \frac{1}{6}   R_{ik0l} x^{k} x^{l} \right)
    \right]  m     \nonumber  \\
    &=&  \mathcal{O} + \mathcal{E} +  
          \left[ \beta 
           \left(1 + a_{i}x^{i} + \frac{1}{2} R_{0k0l} x^{k} x^{l} \right)
        -  \beta \alpha^{i} \left( \omega_{k} \epsilon_{0ijk} x^{j} 
           + \frac{1}{6}   R_{ik0l} x^{k} x^{l} \right)
    \right]  m   \ ,
     \label{HHHH}
\end{eqnarray}
where we have defined $\beta = \gamma^{\hat{0}}$, $\alpha^{i} = \gamma^{\hat{0}} \gamma^{\hat{i}}$ and
$\Pi_{j} = -i \partial_{j} -  eA_{j}$ for brevity.
The even part, $\mathcal{E}$, means that the matrix has only block diagonal elements and 
the odd part, $\mathcal{O}$, means that the matrix has only block off-diagonal elements.
Literally, a product of two even (odd) matrices is even and 
a product of even and odd matrices becomes odd.
In order to take the non-relativistic limit of the Hamiltonian, 
we need to diagonalize the Hamiltonian (\ref{HHHH}) and expand the upper block diagonal part in powers of $1/m$.
More precisely, $1/m$ stands for two dimensionless parameters, $(m x)^{-1}$ and $v/c$.
Here, $x$ represents a typical length scale of 
the system, $v$ is the velocity of the fermion particle and 
$c$ denotes the speed of light.
Assuming $1/m x \ll 1 $ and $v/c \ll 1$, 
which hold in the electron \textit{g}-factor 
measurements~\cite{Hanneke:2008tm,Odom:2006zz},
we will perform the $1/m$ expansion.
This can be carried out both in flat spacetime~\cite{Foldy:1949wa,Bjorken:1965zz}
and in curved spacetime~\cite{Ito:2020wxi}
by repeating unitary transformations order by order in powers of 
$1/m$.
We will follow the procedure shown in~\cite{Ito:2020wxi}.

A unitary transformation to a spinor field is
\begin{equation}
  \psi' = e^{iS} \psi \ ,
\end{equation}
where $S$ is a time-dependent Hermitian 4 $\times$ 4 matrix.
Observing that
%
%
\begin{eqnarray}
  i \frac{\partial \psi'}{\partial t} &=& i \frac{\partial}{\partial t} \left( e^{iS} \psi \right) \nonumber \\
               &=& e^{iS} \left( i \frac{\partial \psi}{\partial t} \right) 
                   + i \left( \frac{\partial}{\partial t} e^{iS} \right) \psi   \nonumber \\
               &=& \left[ e^{iS} H e^{-iS} +  i \left( \frac{\partial}{\partial t} e^{iS} \right)  e^{-iS} \right]
                     \psi'  \ ,
\end{eqnarray}
we see that the Hamiltonian after the unitary transformation is
\begin{equation}
  H' = e^{iS} H e^{-iS} +  i \left( \frac{\partial}{\partial t} e^{iS} \right)  e^{-iS} \ .  \label{traH}
\end{equation}
Taking $S$ to be proportional to powers of $1/m$, 
the transformed Hamiltonian (\ref{traH}) can be expanded
in powers of $S$ up to arbitrary order of $1/m$:
\begin{eqnarray}
  H' &=& H + i \big[ S,H \big] - \frac{1}{2} \big[ S,\big[S,H \big] \big] 
         - \frac{i}{6} \big[S,\big[S, \big[S , H \big] \big] \big]  + \cdots \nonumber \\
     &&  - \dot{S} - \frac{i}{2} \big[ S,\dot{S} \big] 
         + \cdots \ .  \label{newH}
\end{eqnarray}

First, we eliminate the off-diagonal part of the Hamiltonian (\ref{HHHH}) at the order of $m$ by a
unitary transformation.
Then we will drop the higher order terms with respect to 
$a_{i}$, $\omega_{i}$ and the Riemann tensor.
We assume that the time derivative of the Riemann tensor 
is small enough to neglect.
Notice that the time derivative of the spatial coordinate $x^{i}$
is a higher order of $v/c$ and thus we also neglect it%
\footnote{
The Hermiticity of the non-relativistic Hamiltonian 
is guaranteed when the metric is time independent~\cite{Huang:2008kh}.
}.
To cancel the last term in the square bracket of (\ref{HHHH}),
we take
\begin{equation}
  S = - \frac{i}{2m}\beta 
  \left[ - \beta \alpha^{i} \left( \omega_{k} \epsilon_{0ijk} x^{j} 
           + \frac{1}{6}   R_{ik0l} x^{k} x^{l} \right)m \right] \ .
\end{equation}
We then obtain
\begin{eqnarray}
  i \big[ S,H \big] &\simeq&
      \beta \alpha^{i} \left( \omega_{k} \epsilon_{0ijk} x^{j} 
           + \frac{1}{6}   R_{ik0l} x^{k} x^{l} \right) m
     - \frac{1}{2} \big[ \alpha^{i} , \alpha^{j} \big] 
       \left(  \omega_{k} \epsilon_{0ilk} x^{l} 
       + \frac{1}{6} R_{ik0l} x^{k} x^{l} \right) \Pi_{j}
      \nonumber \\
    &&  + \frac{i}{2} \alpha ^{i} \alpha^{j} \left( -\omega_{k} \epsilon_{0jik} 
        + \frac{1}{3} R_{0ikj} x^{k} 
        + \frac{1}{6} R_{0jik} x^{k} \right)
    \ . \label{SH}
\end{eqnarray}
Therefore, from Eqs.\,(\ref{newH}) and (\ref{SH}), we have the transformed Hamiltonian 
with accuracy mentioned above:
%
%
%
%
%
%
%
%
%
%
\begin{eqnarray}
 H' &\simeq& H + i \big[ S,H \big] \nonumber \\
    &=&  \mathcal{O} + \mathcal{E}' +  
          \beta 
           \left(1 + a_{i}x^{i} + \frac{1}{2} R_{0k0l} x^{k} x^{l}  \right) m   \ ,
     \label{HHHHH}
\end{eqnarray}
where we have used the relation $\big\{ \alpha^{i} ,\alpha^{j} \big\} = 2 \delta^{ij}$
and $\mathcal{E}'$ is defined by
\begin{eqnarray}
  \mathcal{E}' = &-& eA_{0}
      + \frac{i}{8} [ \alpha^{i}, \alpha^{j} ] \omega_{k} \epsilon_{0ijk} 
      + \omega_{k} \epsilon_{0ijk} x^{j} \Pi_{i}  \nonumber \\
          &+& \frac{i}{3}   R_{0iki} x^{k}  
          + 
          \frac{2}{3} R_{0kil} x^{k} x^{l} \Pi_{i}
          -\frac{i}{8} [\alpha^{i},\alpha^{j}] 
                        R_{ijk0} x^{k} \ .
\end{eqnarray}
One can see that only even terms remain at the order of $m$, as expected.

Next, let us focus on the order of $m^{0}$ and eliminate the odd terms by a unitary transformation.
In order to do so, we choose the Hermitian operator to be
\begin{equation}
  S' = -\frac{i}{2m} \beta \left[ 
            \mathcal{O}  
            - \alpha^{i} \left( 
            a_{j}x^{j} \Pi_{i} - \frac{i}{2} a_{i} 
            + \frac{1}{2}  R_{0k0l} x^{k} x^{l} \Pi_{i}
            - \frac{i}{2}  R_{0k0i} x^{k} \right)  \right]   \ .
\end{equation}
First of all,
\begin{equation}
  i \big[ S',H' \big] \simeq
    - \mathcal{O} +  i \big[ S',\mathcal{O} \big] +  i \big[ S',\mathcal{E}' \big] \ , 
          \label{S,H}
\end{equation}
Furthermore, up to the order of $1/m$, we find
\begin{equation}
  - \frac{1}{2} \big[ S',\big[S',H' \big] \big]  \simeq
      - \frac{i}{2} \big[ S',  \mathcal{O} \big]  
       \ , 
\end{equation}
and
\begin{equation}
  - \dot{S}' \simeq 
      \frac{i}{2m} \beta  \dot{\mathcal{O}}  
    - \frac{i}{m} \beta \alpha^{i} a_{j} x^{j} e \dot{A}_{i}  \ .
\end{equation}
Therefore, the unitary transformed Hamiltonian is given by
\begin{eqnarray}
  H''  &\simeq& H' + i \big[ S',H' \big] - \frac{1}{2} \big[ S',\big[S',H' \big] \big] - \dot{S}'  \nonumber \\
  &\simeq& H' 
      + \frac{i}{2} \big[ S',\mathcal{O} \big]
      + i \big[ S',\mathcal{E}' \big]
      - \dot{S}'  \nonumber \\
  &\simeq&  
      \frac{i}{m} \alpha^{i} a_{j} x^{j} e E_{i}
    - \frac{i}{4m} \beta \alpha^{j} R_{0k0l} x^{k} x^{l} e E_{j}
    + \frac{1}{2m} \beta \left(  \big[ \mathcal{O}, \mathcal{E}' \big] 
    + i \dot{\mathcal{O}} \right) \nonumber \\
  && + \mathcal{E}'
     + \frac{1}{2m} \beta \mathcal{O}^{2} 
     + \frac{i}{2m} \beta \left( a^{i}  +  R_{0k0i} x^{k} \right) \Pi_{i}
     + \frac{i}{8m} \beta \big[ \alpha^{i}, \alpha^{j} \big] \left( a^{i} +
                                R_{0k0i} x^{k} \right) \Pi_{j} 
       \nonumber \\
  && 
     - \frac{1}{2m} \beta \alpha^{i} \alpha^{j} \left( a_{k} x^{k} 
     + \frac{1}{2} R_{0k0l} x^{k} x^{l}  \right) 
       \Pi_{i} \Pi_{j} 
     + \frac{1}{8m} \beta  R_{0i0i}
            \nonumber \\
  && + \beta  \left(1 + a_{i}x^{i} + \frac{1}{2} R_{0k0l} x^{k} x^{l} \right) m   
     \nonumber \\
  &=&  \mathcal{O}' + \mathcal{E}'' 
     + \beta  \left(1 + a_{i}x^{i} + \frac{1}{2} R_{0k0l} x^{k} x^{l} \right) m \ ,
\end{eqnarray}
where $E_{j} \equiv \partial_{j} A_{0} -  \dot{A}_{j}$ is an electric field.
We see that $\mathcal{O}'$ consists of only terms of the order of $1/m$, so that 
odd terms at the order of $m^{0}$ have been eliminated correctly.

Finally, again,
we can eliminate the odd term $\mathcal{O}'$ by an appropriate 
unitary transformation. 
The resultant Hamiltonian consists of only even terms up to the order of $1/m$, 
which we want to get.
Thus, up to the order of $1/m$, we have
\begin{equation}
  H'''   \simeq  \mathcal{E}'' 
      + \beta \left(1 + a_{i}x^{i} + \frac{1}{2} R_{0k0l} x^{k} x^{l} \right)
        m \ , \label{H'''}
\end{equation}
where $\mathcal{E}''$ is 
\begin{eqnarray}
   \mathcal{E}'' &=& 
     - eA_{0}   
     - \frac{i}{8} [ \alpha^{i}, \alpha^{j} ] \omega_{k} \epsilon_{0ijk} 
      - \omega_{k} \epsilon_{0ijk} x^{j} \Pi_{i} 
          + \frac{i}{3}   R_{0iki} x^{k}  
          + \frac{2}{3} R_{0kil} x^{k} x^{l} \Pi_{i}
          -\frac{i}{8} [\alpha^{i},\alpha^{j}] R_{ijk0} x^{k}
            \nonumber \\
  &&   + \frac{1}{2m} \beta \mathcal{O}^{2} 
     + \frac{i}{2m} \beta \left( a^{i}  +  R_{0k0i} x^{k} \right) \Pi_{i}
     + \frac{i}{8m} \beta \big[ \alpha^{i}, \alpha^{j} \big] \left( a^{i} +
                                R_{0k0i} x^{k} \right) \Pi_{j} 
       \nonumber \\
  && 
     - \frac{1}{2m} \beta \alpha^{i} \alpha^{j} \left( a_{k} x^{k} 
     + \frac{1}{2} R_{0k0l} x^{k} x^{l}  \right) 
       \Pi_{i} \Pi_{j} + \frac{1}{8m} \beta  R_{0i0i}  \label{E''}  \ .
\end{eqnarray}
Moreover, the first term in the second line of Eq.\,(\ref{E''}) can be evaluated as
\begin{eqnarray}
  \frac{1}{2m} \beta \mathcal{O}^{2} &\simeq&
       \frac{1}{2m} \beta 
       \Big( \delta_{ij} \left( 1 + 2 a_{k}x^{k} \right) + 2 \theta_{ij} \Big)
       \Pi_{i} \Pi_{j}
     + \frac{i}{8m}\beta  \big[ \alpha^{i} ,\alpha^{j} \big] \epsilon_{0ilm} e B^{m}
     \Big(  \delta_{lj}(1 + 2 a_{k}x^{k})  + 2  \theta_{lj}  \Big) \nonumber \\
     && -\frac{i}{4m}\beta \big[ \alpha^{i} ,\alpha^{j} \big]
     \left(  a_{i} \Pi_{j} +
     \frac{1}{4} R_{lmji}  +  \delta^{l}_{j}R_{0i0m} \right) x^{m} \Pi_{l} 
     + \frac{i}{12 m}\beta R_{kikj} x^{j} \Pi_{i} 
     -\frac{i}{m}\beta R_{0i0j} x^{i} \Pi_{j}
     \nonumber \\
   &&  
      - \frac{1}{4m} \beta R_{0i0i}  
      - \frac{1}{16m} \beta \alpha^{i} \alpha^{j} \alpha^{k} \alpha^{l} R_{ijkl}
      + \frac{i}{16m} \beta \big\{ \alpha^{i} , \alpha^{j} \alpha^{k} \alpha^{l} \big\}
        R_{kljm} x^{m} \Pi_{i}  \ ,
    \label{Ono2}
\end{eqnarray}
where $B^{i} \equiv \frac{1}{2} \epsilon_{0ijk} (\partial_{j}A_{k} - \partial_{k}A_{j})$
is a magnetic field%
\footnote{In general, an external magnetic field itself would be modified by inertial and gravitational effects 
as was explicitly shown for a simple system like a Hydrogen atom~\cite{Parker:1980hlc,Parker:1980kw,Perche:2020lzz}.
We ignore such corrections since there is no way to evaluate them model-independently, namely
they depend on detail of a mechanism for creating an external magnetic field.
}.
Using Eqs.\,(\ref{E''}), (\ref{Ono2}) and the relation, 
$\big[ \alpha^{i} ,\alpha^{j} \big] = 2 i \epsilon_{0ijk} \sigma^{k}$,
in the transformed Hamiltonian (\ref{H'''}),
we finally arrive at the Hamiltonian for a non-relativistic fermion/anti-fermion up to the order of $1/m$:
\begin{eqnarray}
  H'''  &=&   \beta \left( 1 + a_{i}x^{i} + \frac{1}{2} R_{0k0l} x^{k} x^{l} \right)   m  \nonumber \\
  && - eA_{0}   
     - \omega_{k} \epsilon_{0ijk} x^{i} \Pi_{j}  - \omega_{i} S^{i}
     + \frac{1}{3}  R_{0kil} \left( \Pi_{i} x^{k} x^{l}  
     +  x^{k} x^{l}  \Pi_{i} \right)  
     + \frac{1}{2} \epsilon_{0ijl} S^{l} R_{ijk0}  x^{k} \nonumber \\
  && - \frac{e}{m}\beta S^{i} B^{j} \left[  \delta_{ij} 
       \left( 1 + a_{k} x^{k} +
        \frac{1}{2} R_{0k0l} x^{k} x^{l} + \frac{1}{6} R_{mkml} x^{k} x^{l}  \right)
              - \frac{1}{6}  R_{ikjl} x^{k} x^{l}  \right]   \nonumber \\
   && + \frac{1}{2m}\beta \Pi_{i} 
            \left[ \delta_{ij} \left( 1 + a_{k} x^{k} + \frac{1}{2} R_{0k0l} x^{k}x^{l} \right)
              + \frac{1}{3}  R_{jkil}  x^{k} x^{l} \right] \Pi_{j}  \nonumber \\
   && +\frac{1}{2m}\beta \epsilon_{0ijk} S^{k}  a^{i} \Pi^{j} 
      + \frac{1}{4m}\beta \epsilon_{ijk} S^{k} 
       \left(  R_{ijlm}  + 2 \delta_{jm} R_{0i0l} \right) x^{l} \Pi_{m}  \nonumber \\
   && 
      - \frac{1}{8m} \beta R_{0i0i}  + \frac{1}{24m} \beta R_{ijij} \ ,
      \label{Hddd}
\end{eqnarray}
where we have replaced the canonical momentum in flat spacetime by one in curved spacetime~\cite{2013PhRvA..88f2117G}:
$\Pi_{i} \rightarrow  \Pi_{i} + i \frac{1}{\sqrt{-g}} \partial_{i} \sqrt{-g}  = \Pi_{i} - \frac{i}{6} R_{jijk} x^{k}$
($g$ is the determinant of the metric).
Also a spin has been defined by $S^{i} = \sigma^{i}/2$ with the Pauli matrices $\sigma^{i}$.
Note that if one wants to consider an anti-fermion in the above Hamiltonian,  
one needs to take charge conjugation for the wave function of the lower two compoents.
The terms for inertial effects coincide with an earlier 
work~\cite{Singh:2000xq}.
On the other hand, several terms for gravitational effects are updated 
compared with the previous work~\cite{Ito:2020wxi} where there was a miscalculation. 
The first parenthesis represents the rest mass and its corrections due to
inertial and gravitational effects.
The inertial one is recognized as an usual inertial force 
and the gravitational one is the leading order gravitational modification
to a particle trajectory as we will see later.
The third term~\cite{Werner:1979gi} corresponds to the Coriolis force.
The fourth term is the spin-rotation coupling~\cite{Mashhoon:1988zz},
which modifies the magnetic moment, i.e., \textit{g}-factor.
In the fifth term, we can clearly see that a correct operater ordering has been derived automatically.
Interestingly, the sixth term represents the effect of a dipole structure, namely the spin angular momentum, on the 
trajectry of a freely falling particle in curved spacetime as can be seen in the 
Mathisson-Papapetrou-Dixon equation\footnote{
Strictly speaking, we should have taken the term into account when we constructed a proper reference frame.
Indeed, the effect can be treated as a linear acceleration of a paticle and can be included 
in $a_{i}$ in the metric (\ref{met033}).
Then the sixth term disappers.}.
The third line represents corrections of the magnetic moment 
due to gravity.
The fourth line shows that also the kinetic term is modified by 
inertial and gravitational effects.
In the fifth line, we find
the inertial spin-orbit coupling~\cite{Hehl:1990nf}
and the gravitational spin-orbit coupling~\cite{Ito:2020wxi}, respectively.
The sixth line consists of energy shifts due to gravity
at the order of $1/m$.
\section{Particle trajectory and spin kinematics in gravity} 
\label{eacheffect}
In this section, we 
investigate the inertial and the gravitational effects in 
the Hamiltonian (\ref{Hddd})
to trajectories and spin kinematics of a Dirac particle in the presence of
an external magnetic field.
In the sections \ref{PT} and \ref{SK}, we will show that 
the cyclotron and the Larmor frequencies are corrected according to modifications of 
particle trajectories and spin kinematics, respectively.
In the section \ref{SO}, it will turn out that the spin-orbit couplings modify the cyclotron and the Larmor frequencies simultaneously.

\subsection{Particle trajectries}\label{PT}
In terms of the canonical momentum $p_{j} = \Pi_{j} + eA_{j}$,
a part concerned with particle trajectories in the Hamiltonian (\ref{Hddd}) is
\begin{eqnarray}
  H_{{\rm orbit}} &=& 
   \left( 1 + a_{i}x^{i} + \frac{1}{2} R_{0k0l} x^{k} x^{l} \right)   m  
     - eA_{0} - \omega_{k} \epsilon_{0ijk} x^{i} 
       \left( p_{j} -  eA_{j} \right)   
        \nonumber \\
 &&  + \frac{1}{2m} 
            \left[ \delta_{ij} \left( 1 + a_{k} x^{k} + \frac{1}{2} R_{0k0l} x^{k}x^{l} \right)
              + \frac{1}{3}  R_{jkil}  x^{k} x^{l} \right] 
              \left( p_{i} -  eA_{i} \right) \left( p_{j} -  eA_{j} \right)
       \ ,
  \label{Horbit}
\end{eqnarray}
where we have neglected higher order terms with respect to 
$(m x)^{-1}$ and $v/c$ in several parts.
From the Hamiltonian, one can derive the classical
equation of motion for a particle 
in the presence of an external magnetic field:
\begin{equation}
  \ddot{x}^{i} = 
    - a^{i} 
    - R_{0i0j} x^{j} 
  + \left[ 
    \left\{ \delta_{ij} 
    \left( 1 + a_{m}x^{m} + \frac{1}{2} R_{0m0n} x^{m}x^{n} \right)
    + \frac{1}{3} R_{imjn} x^{m}x^{n}  \right\}
    \frac{e}{m}B^{l}
      + 2 \delta_{ij} \omega^{l} 
        \right] \epsilon_{0jkl} \dot{x}^{k}  \ , \label{eqmo}
\end{equation}
where a dot denotes a derivative with respect to $x^{0}$.
Note that we neglected an external electric field, which should exist
in actual Penning trap experiments (see the appendix \ref{review}) 
because it is unnecessary to examine effects of gravity up to
the order of $1/m$.  
Because of the magnetic field,
the particle experiences the cyclotron motion with the 
frequency $2\pi f_{c} = eB/m$. 
The square bracket in Eq.\,(\ref{eqmo}) shows that 
the frequency is directly modified by
the inertial and the gravitational effects.
The second term of the right-hand side in Eq.\,(\ref{eqmo})
also modifies the cyclotron frequency as we will see soon.

For concreteness, let us consider the gravitational potential of the Earth,
$\phi = - G \frac{M}{|\bm{x} - \bm{x}_{0}|}$, as the source of 
the curvature.
$G$ is the gravitational constant, $M$ is the mass of the Earth and
$\bm{x}_{0}$ denotes the center of the Earth.
Then a component of the Riemann tensor which is evaluated 
at $\bm{x}=0$ is 
\begin{equation}
  R_{0i0j} = \phi_{,ij} =
  - 3G\frac{M}{x_{0}^{5}} x_{0}^{i} x_{0}^{j}
  + G\frac{M}{x_{0}^{3}} \delta_{ij} \ .
\end{equation}
This indicates that 
$R_{0i0j}x^{i}x^{j} \ll a_{i}x^{i} 
= - G\frac{M}{x_{0}^{3}} x_{0}^{i}x^{i}$.
Therefore the curvature tensors in the square bracket are negligible compared 
with the inertial effects.
Then, the corrections 
to the cyclotron frequency  
in the square brackets of Eq.\,(\ref{eqmo}) can be estimated as
\begin{equation}
  \tilde{f}_{c} \simeq f_{c} 
    \left( 1 + a_{i}x_{cy}^{i} \pm \frac{2\omega}{2\pi f_{c}} \cos\theta  \right) \ ,
    \label{tilde}
\end{equation}
where $\theta$ is an angle between $B^{i}$ and $\omega^{i}$,
and $x_{cy}$ represents a position vector for the cyclotron motion. 
At the third term, 
the upper sign is for a positive charge fermion 
(negative charge anti-fermion) and 
the lower sign is for a negative charge fermion 
(positive charge anti-fermion).

Next, we evaluate the second term of the right-hand side in Eq.\,(\ref{eqmo}).
To this end, we consider the equation:
\begin{eqnarray}
  \ddot{x}^{i} &=& 
  - a^{i}
  - R_{0i0j} x^{j} 
  +  \frac{e}{m}B^{l}
      \epsilon_{0ikl} \dot{x}^{k}  \nonumber \\
    &=& 
  - a^{i}
  + \left[ 
    3G\frac{M}{x_{0}^{5}}  
     x^{i}_{0} x^{j}_{0} x^{j}
  - G\frac{M}{x_{0}^{3}} x^{i}
  \right]
  +  \frac{e}{m}B^{l}
      \epsilon_{0ikl} \dot{x}^{k}
      \ . \label{eqmo2}
\end{eqnarray}
We now take the direction of the magnetic field to be the $z$-direction
$\bm{B} = (0,0,B^{z})$ and 
assume that $\bm{x}_{0}=(0,0,z_{0})$, namely 
the magnetic field is perpendicular to the Earth's surface.
Then the cyclotron orbit is on the $x$-$y$ plane and the equations of motion 
are
\begin{subequations}
\begin{empheq}[left=\empheqlbrace]{align}
  \ddot{x} &=  - G\frac{M}{x_{0}^{3}} x
                   + \frac{eB^{z}}{m} \dot{y}  \ , \label{xeqmo} \\
  \ddot{y} &=  - G\frac{M}{x_{0}^{3}} y
                   - \frac{eB^{z}}{m} \dot{x}  \ , \label{yeqmo} \\ 
  \ddot{z} &= - a^{z} + 2G\frac{M}{x_{0}^{3}} z \ . \label{zeqmo}
\end{empheq}
\end{subequations}
In Eq.\,(\ref{zeqmo}), the first term is for the free fall motion.
The second term is a tidal effect and
modifies the axial frequency $f_{z}$ explained 
in the appendix \ref{review}.
However, it is not relevant because the observed $g$-factor
(\ref{gfac}) is not affected by modulation of $f_{z}$.
Eqs.\,(\ref{xeqmo}) and (\ref{yeqmo}) represent the cyclotron motion
with a gravitational modification and they can be solved as 
\begin{subequations}
\begin{empheq}[left=\empheqlbrace]{align}
    x & =  C_{1} \cos (-2\pi f_{+}t) + C_{2} \cos(-2\pi f_{-}t)   \ ,  \\
    y & =  C_{1} \sin(-2\pi f_{+}t) + C_{2} \sin(-2\pi f_{-}t)      \ ,
\end{empheq}
\end{subequations}
where $C_{1}$ and $C_{2}$ are integration constants and
\begin{equation}
  2\pi f_{\pm} = \frac{2\pi f_{c} \pm 
  \sqrt{ (2\pi f_{c})^{2} +  4GM/x_{0}^{3}}}{
         2  } \ .  \label{pm}
\end{equation}
A modified cyclotron frequency should be the plus sign and it can be approximated as
\begin{equation}
  f_{+} \simeq f_{c} + 
    \frac{ GM / x_{0}^{3}}{(2\pi f_{c})^{2}} \ .
\end{equation}
Together with Eq.\,(\ref{tilde}), we find that 
the total modification to the cyclotron frequency is
\begin{equation}
  \bar{f}_{c}^{(A)} 
     =
  f_{c} \left( 1 + a_{i}x_{cy}^{i} 
              \pm \frac{2\omega}{2\pi f_{c}} \cos\theta 
                    + \frac{ GM / x_{0}^{3}}{(2\pi f_{c})^{2}} 
                    \right) \ . \label{cymodi}
\end{equation}
\subsection{Spin kinematics}\label{SK}
In the Hamiltonian (\ref{Hddd}),
the dynamics of a spin in the presence of the magnetic field and gravity is 
determined by 
\begin{eqnarray}
  H_{{\rm spin}} &=&
     \frac{1}{2} \epsilon_{0ijl} S^{l} R_{ijk0} x^{k}
    - \omega_{i} S^{i} \nonumber \\
    &&- \frac{e}{m} S^{i} B^{j} \left[  \delta_{ij} 
       \left( 1 + a_{k} x^{k} +
        \frac{1}{2} R_{0k0l} x^{k} x^{l} + \frac{1}{6} R_{mkml} x^{k} x^{l}  \right)
              - \frac{1}{6}  R_{ikjl} x^{k} x^{l}  \right] 
       \ .
  \label{Hspin}
\end{eqnarray}
Note that the spin-orbit couplings will be treated in the next subsection independently.
As discussed in the previous subsection, 
the contribution from the curvature terms is negligible compared with 
the inertial effects.
Moreover, the component of the curvature, $R_{ijk0}$, is zero 
for the gravitational potential of the Earth.
Then, from the Hamiltonian Eq.\,(\ref{Hspin}), one can derive the 
Heisenberg equation of motion:
\begin{equation}
  \dot{S}^{a} =  - \epsilon_{0aib} S^{b} 
     \left[ \frac{e}{m}B^{i} \left( 1 + a_{k}x^{k} \right) 
            + \omega_{i} \right] \ .
\end{equation}
It shows that 
the spin precession is induced by the external magnetic field 
with the Larmor frequency, $2\pi f_{{s}} = eB/m$, but
the frequency is modified by the inertial effects.
Notice that we do not consider loop corrections to the magnetic moment, i.e.,
$g$-factor is replaced by $2$.
The modified Larmor frequency is estimated as
\begin{equation}
  \bar{f}_{s}^{(B)} = f_{s} \left( 1 + a_{i} x_{cy}^{i} 
           \pm \frac{\omega}{2\pi f_{s}} \cos\theta   \right) \ .
           \label{spimodi}
\end{equation}
At the third term, again,
the upper sign is for a positive charge fermion 
(negative charge anti-fermion) and 
the lower sign is for a negative charge fermion 
(positive charge anti-fermion).
\subsection{Spin-orbit coupling}\label{SO}
\begin{figure}[h]
\centering
\includegraphics[width=10.0cm]{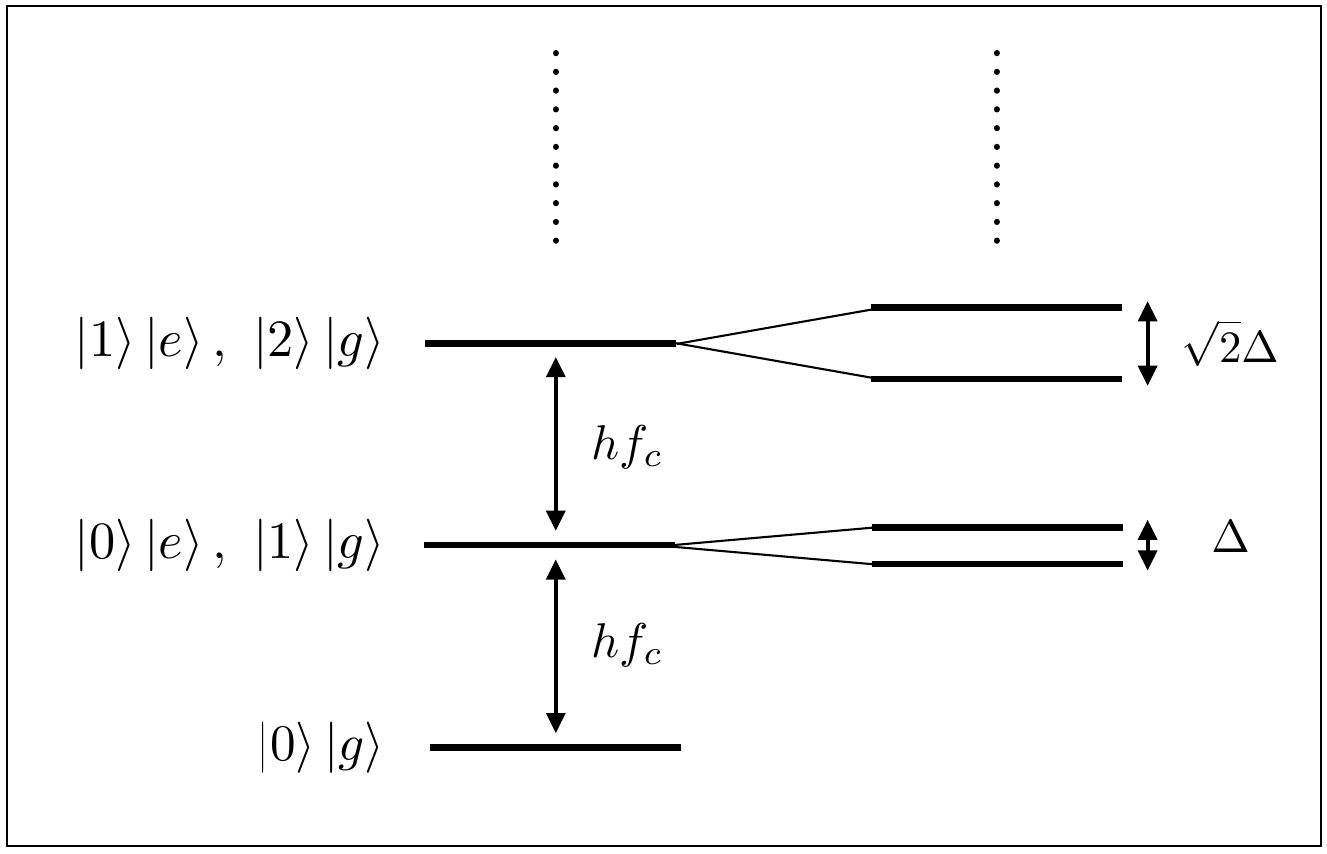}
\caption{The diagonalized energy levels are depicted.
The degenerated states, $\ket{n}\ket{e}$ and $\ket{n+1}\ket{g}$, are split 
by $\Delta\sqrt{n+1}$ due to the spin-orbit coupling.}
\label{split}
\end{figure}

So far, we have studied how the cyclotron and the Larmor frequencies are 
modified by gravity individually.
However there are the inertial and the gravitational 
spin-orbit couplings, those are
\begin{equation}
  H_{{\rm spin-orbit}} =
     \frac{1}{2m}  \Pi_{i} \Pi_{i}
     - \frac{e}{m} S^{i} B^{i}
     + \frac{1}{2m} \epsilon_{0ijk} S^{k}  a^{i} \Pi^{j} 
     + \frac{1}{4m} \epsilon_{0ijk} S^{k} 
       \left(  R_{ijlm}  + 2 \delta_{jm} R_{0i0l} \right) x^{l} \Pi_{m}
       \ , \ 
  \label{Hspinorbit}
\end{equation}
where we have incorporated the free parts of 
the kinetic and the spin precession terms.
The third and the fourth terms stand for the inertial and the gravitational 
spin-orbit couplings, respectively.
They would make the energy levels split as in the case 
of a Hydrogen atom.
Let us investigate the energy split in details.
First of all, as mentioned in the previous subsections, 
the gravitational spin-orbit coupling terms should be smaller
than the inertial one as long as we consider the Earth as a source of the curvature.
Thus, we neglect them.
When we set the magnetic field to be along with the $z$-direction,
$\bm{B} = (0,0,B^{z})$, the Hamiltonian can be rewritten as follows:
\begin{equation}
  H_{{\rm spin-orbit}} =
     (2\pi f_{c}) \left( \alpha^{\dagger} \alpha + \frac{1}{2}  \right) 
     - (2\pi f_{s}) S^{z}
     - \Delta
       \left( \alpha S_{+}
              + \alpha^{\dagger} S_{-} \right)
       \ .
  \label{Hspinorbit2}
\end{equation}
Here again, we have set $\bm{a} = (0,0,a_{z})$ by assuming that the 
magnetic field is perpendicular to the Earth's surface.
The coupling constant 
\begin{equation}
\Delta=\sqrt{\frac{2\pi f_{c}}{8m}} a \ ,  \label{Delta}
\end{equation}
has been defined for brevity.
We also defined creation and annihilation operators for 
the cyclotron motion,
\begin{equation}
  \alpha = \frac{1}{\sqrt{2eB^{z}}} \left( -i \Pi_{x} + \Pi_{y} \right) \ , \quad
  \alpha^{\dagger} = \frac{1}{\sqrt{2eB^{z}}} \left( i \Pi_{x} + \Pi_{y} \right) \ ,
\end{equation}
($[\alpha , \  \alpha^{\dagger}] = 1$) and ladder operators for the spin,
\begin{equation}
  S_{+} = S^{x} - i S^{y} \ , \quad   S_{-} = S^{x} + i S^{y} \ .
\end{equation}
The Hamiltonian (\ref{Hspinorbit2}) is nothing but the 
Jaynes-Cummings model~\cite{1443594}.
We have implicitly set, $2\pi f_{c} = 2\pi f_{s} = eB/m$, in 
Eq.\,(\ref{Hspinorbit2}) since
the inertial spin-orbit interaction is our sole concern now and other corrections are negligible, at least, at the linear order.
Without the spin-orbit coupling in the Hamiltonian (\ref{Hspinorbit2}),
the eigenstates are specified by
$\ket{n}\ket{g}$ and $\ket{n}\ket{e}$, where 
$n$ is an eigenvalue of the number operator $\alpha^{\dagger} \alpha$ and
$\ket{g}$ ($\ket{e}$) represents the ground (excited) state for the spin.
Then we find that the two states, $\ket{n}\ket{e}$ and $\ket{n+1}\ket{g}$
are degenerates.
However, in fact, this degeneracy is resolved 
due to the presence of the spin-orbit coupling.
Diagonalizing the Hamiltonian (\ref{Hspinorbit2}) in the subspace spanned by
$\ket{n}\ket{e}$ and $\ket{n+1}\ket{g}$, 
one can obtain the split energy levels 
%
%
%
\begin{equation}
  E_{n} = 2\pi f_{c} \left( n + 1 \right) \pm  
          \frac{1}{2} \Delta \sqrt{ \left( n+1 \right) } \label{enelevel}
          \ .
\end{equation}
Eq.\,(\ref{enelevel}) shows that the each pair of the degenerated states is split by
$\Delta\sqrt{n+1}$.
Therefore if we observe energy transitions for larger $n$,
the energy split becomes larger.
The diagonalized energy levels are depicted in Fig.\,\ref{split}.
In the case of the Penning trap experiments,
we observe one quantum transition from the ground state $\ket{0}\ket{g}$.
Therefore,
following dressed cyclotron and Larmor frequencies are detected,
\begin{equation}
  \bar{f}_{c}^{(C)} = f_{c}\left( 1 
       - \frac{1}{4\sqrt{2}}
         \frac{a}{\sqrt{(2 \pi f_{c})m}} \right) \ ,
  \quad
  \bar{f}_{s}^{(C)} = f_{s}\left( 1 
       + \frac{1}{4\sqrt{2}}
         \frac{a}{\sqrt{(2 \pi f_{s})m}} \right) \ .
  \label{somodi}
\end{equation}
\section{Detectability in electron $g$-factor measurements} 
\label{detectability}
In this section, 
we reveal how the general relativistic corrections investigated 
in the previous section appear in the observed $g$-factor (\ref{gfac}) in
Penning trap experiments.
Furthermore, we estimate magnitude of the 
corrections and discuss its detectability
in a concrete case of the electron $g$-factor 
measurement~\cite{Hanneke:2008tm}.

As is discussed in the appendix \ref{review},
the observed $g$-factor 
in Penning trap experiments is Eq.\,(\ref{gfac}).
On the other hand, we revealed the general relativistic corrections on
the cyclotron and the Larmor frequencies in the 
previous sections, from Eqs.\,(\ref{cymodi}), (\ref{spimodi}) and (\ref{somodi}),
that is
 \begin{empheq}[left=\empheqlbrace]{align}
    \label{51}
    \bar{f}_{c} & = f_{c} \left( 1 + a_{i}x_{cy}^{i} 
    \pm \frac{2\omega}{2\pi f_{c}} \cos\theta 
    - \frac{1}{4\sqrt{2}}
         \frac{a}{\sqrt{(2 \pi f_{c})m}}                   
    + \frac{ GM / x_{0}^{3}}{(2\pi f_{c})^{2}} 
                    \right) \ ,    \\
    \label{52}
    \bar{f}_{s} & = f_{s} \left( 1 + a_{i} x_{cy}^{i} 
    \pm \frac{\omega}{2\pi f_{s}} \cos\theta   
    + \frac{1}{4\sqrt{2}}
         \frac{a}{\sqrt{(2 \pi f_{c})m}}  \right)    \ .
 \end{empheq}
%
%
%
From the above equations, general relativistic corrections on
the observed $g$-factor can be read:
\begin{eqnarray}
  \frac{\delta g}{2} &=&  \frac{\bar{f}_{s}}{\bar{f}_{c}} 
                       - \frac{f_{s}}{f_{c}}  \nonumber \\
                     &\simeq&
   \mp  \frac{\omega}{2\pi f_{c}} \cos\theta 
   + \frac{1}{2\sqrt{2}}
         \frac{a}{\sqrt{(2 \pi f_{c})m}} 
   - \frac{ GM / x_{0}^{3}}{(2\pi f_{c})^{2}} \ .   \label{modimodi}
   \label{deltag}       
\end{eqnarray}
One can find that the second terms in the right-hand side of 
Eqs.\,(\ref{51}) and (\ref{52}) canceled out.
Each term in Eq.\,(\ref{modimodi}) has different dependence on $f_{c}$ and $m$.
Therefore, their magnitude change depending on situations.

We now estimate the magnitude of the general relativistic corrections
in Eq.\,(\ref{deltag}), 
in particular for~\cite{Hanneke:2008tm}.
The experiment was conducted in
Harvard University whose longitude is $42.4^{\circ}$.
Thus, the angle between the Earth's rotation vector $\bm{\omega}$ and
the magnetic field $\bm{B}$ 
which is assumed to be perpendicular to the surface of the Earth would be
$\theta \simeq 0.674 \, {\rm rad}$.
Furthermore, using values, $\omega = 7.27 \times 10^{-5} \, {\rm rad/s}$, 
$a = 9.81 \, {\rm m/s}^{2}$,
$G = 6.67 \times 10^{-11}\,  {\rm m^{3}/kg \, s^{2}}$,
$M = 5.97 \times 10^{24} \, {\rm kg}$, 
$x_{0} = 6.38 \times 10^{6} \, {\rm m}$,
$m = 511 \, {\rm keV}$ 
and 
$f_{c} = f_{s} = eB/m \simeq 150 \, {\rm GHz}$ (see~\cite{Hanneke:2008tm}),
we can estimate each correction.
The result is summarized in Table~\ref{electron}.
From Table~\ref{electron}, we see that the correction from the tidal effect is much smaller than other effects of inertial ones
as expected.
It should be mentioned that the correction from the the gravity of Earth, $\bm{a}$, is much larger than the previous report~\cite{Ulbricht:2019dzm} where effects of $\bm{a}$ on the electron $g$-factor was studied.
It is because that they investigated a correction from non linear contribution of $\bm{a}$ but 
did not focus on the spin-orbit coupling induced by $\bm{a}$, which is linear order contribution.
Table~\ref{electron} shows that the effects of Earth's rotation cause
the most largest correction to the electron $g$-factor%
\footnote{In~\cite{Notari:2019qcx} where general relativistic corrections to the muon $g$-factor is mainly studied, 
rough estimation of effects of the Earth's rotation on the electron $g$-factor is given.
Although that estimation is one order of magnitude bigger than our result, we believe that the discrepancy largely comes from a missing factor of $1/2\pi$ in their calculation.}
in the case of~\cite{Hanneke:2008tm}.
It can be detected if the current uncertainty
$\Delta g/2 \simeq 2.8 \times 10^{-13}$~\cite{Hanneke:2008tm} is improved 
by 4 orders of magnitude.
Therefore it would be important to consider the effects of gravity for 
future more accurate experiments~\cite{Gabrielse:2019cgf,Fan:2020enk}.
\begin{table}[h]
\centering
\includegraphics[width=13.6cm]{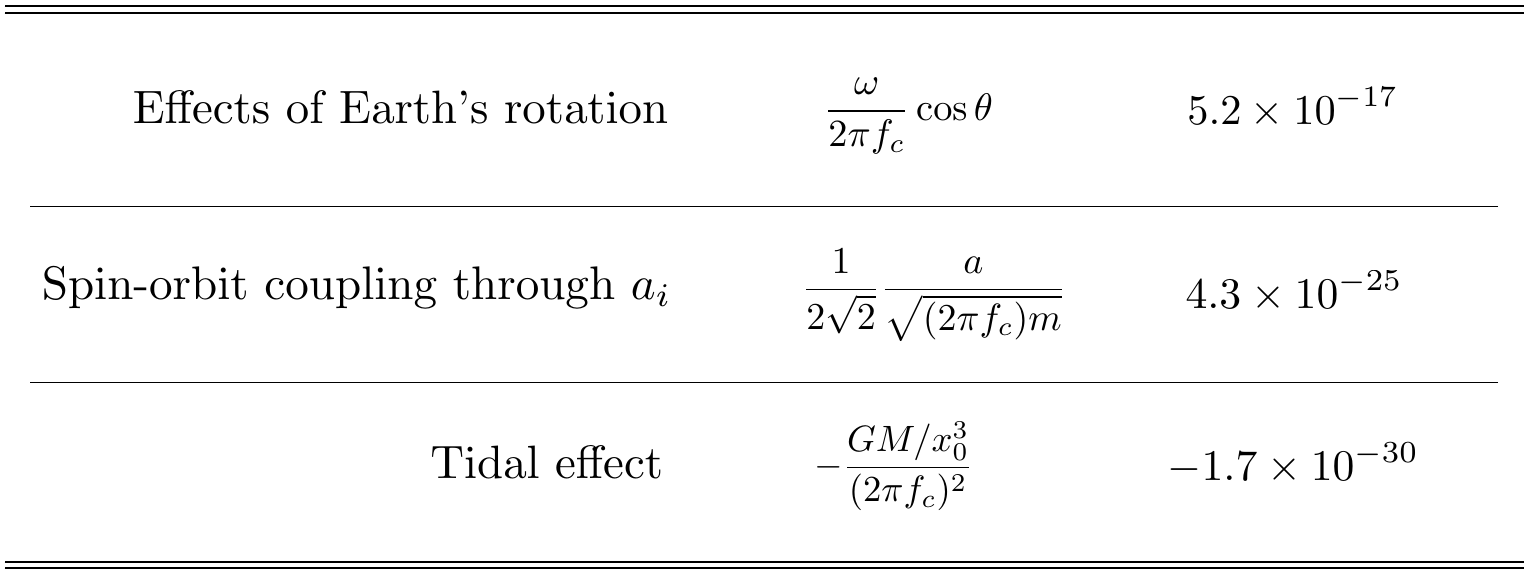}
\caption{The general relativistic corrections in
Eq.\,(\ref{deltag}) for the case of the electron $g$-factor 
measurement~\cite{Hanneke:2008tm} are listed.
}
\label{electron}
\end{table}
\section{Conclusion}
Electron $g$-factor measurements have been operated with remarkable high 
accuracy~\cite{Hanneke:2008tm,Odom:2006zz,VanDyck:1987ay}.
Effects of gravity may not be negligible at the current sensitivity, 
so that quantitative and comprehensive study of general relativistic
corrections in $g$-factor measurements is desired.
In the first part of this paper, we revealed all linear order
inertial and gravitational effects on a Dirac particle
up to the order of $1/m$ in a proper reference frame, which is represented by the Hamiltonian (\ref{Hddd}).
The Hamiltonian (\ref{Hddd}) is useful to investigate gravitational and inertial effects on any systems consistent 
with approximations we have made as it partly has been done for the case of a Hydrogen atom~\cite{Parker:1980hlc,Parker:1980kw,Perche:2020lzz}%
\footnote{\cite{Perche:2020lzz} appeared after
submission of our paper.}.
In the later part of this paper, we applied the Hamiltonian (\ref{Hddd}) to 
Penning trap experiments where a Dirac particle experiences 
the cyclotron motion and the spin precession in a cavity, 
i.e., a geonium atom, and evaluated 
the magnitude of the effects of gravity.

It turned out that gravity modifies the cyclotron motion and 
the spin precession in various ways.
These effects were investigated in the section \ref{eacheffect} in details.
Importantly, each general relativistic correction
has different dependence on the cyclotron frequency
$f_{c} = eB/m$ and the mass $m$.
Therefore, the magnitude of each contribution can differ 
in situations.

In the section \ref{detectability}, 
we considered an electron $g$-factor measurement and 
estimated the magnitude of each correction.
The result is summarized in Table~\ref{electron}.
The most largest correction comes from the effects of the rotation of
the Earth and it can be detected if the current sensitivity is improved
by 4 orders of magnitude.
Therefore it would be important to consider the effects of gravity for 
future more accurate experiments~\cite{Gabrielse:2019cgf,Fan:2020enk}.

Finally, we mention that
our discussion can be applied to cases for $g$-factor measurements of 
positron~\cite{Gabrielse:2019cgf,VanDyck:1987ay,Schwinberg:1981ev},
the proton~\cite{Rodegheri:2012zz,DiSciacca:2012uz} and 
the antiproton~\cite{DiSciacca:2013hya,Gabrielse:1999kc} in parallel.
It is explicitly shown in Eq.\,(\ref{deltag}) that 
the proton and the antiproton (the electron and the positron) 
should take the minus (plus) sign.
For the case of the muon~\cite{Muong-2:2021ojo,Bennett:2006fi,Bailey:1978mn}, 
we need a more careful consideration because the velocity of the muon
is in special relativistic regime.
In the present paper, we have focused on general relativistic effects that 
are leading order with respect to $v/c$ and neglected higher order terms.
Therefore, the accuracy of the approximation would be worse if we
apply the discussion to the case of the muon.
However, of course, since $v/c < 1$ is satisfied even for the case of 
the muon, our result still would be valid to estimate magnitude of general 
relativistic corrections in the muon $g$-factor measurements.
For instense, we can evaluate the correction from the Earth's rotation to the 
muon $g$-factor measurement conducted at the Fermi National Accelerator Laboratory~\cite{Muong-2:2021ojo} as
$\delta g/2 \sim 1.2 \times 10^{-12}$, which coincides with the estimation in~\cite{Notari:2019qcx}.
Compared with the electron case, the correction is relatively colose to the current uncertainty of the muon $g$-factor 
measurement $\Delta g / 2 \sim 5.4 \times 10^{-10}$~\cite{Muong-2:2021ojo}.
Therefore, it may also be detectable in the near future.
\begin{acknowledgments}
A.\,I\,. was supported by JSPS KAKENHI Grant Numbers JP17H02894, JP17K18778 and 
National Center for Theoretical Sciences.
\end{acknowledgments}
\appendix 
\section{Brief review of Penning trap experiments}
\label{review}
In this appendix, we give an overview of Penning trap experiments and 
identify an observable in electron $g$-factor measurements.
More detailed discussions would be found 
in~\cite{Brown:1985rh,DUrso:2003ilf,Hanneke:2010au}.

First of all, to measure the electron $g$-factor, 
we apply an external magnetic field on an electron.
Then, the electron experiences the cyclotron motion and the spin precession with 
the cyclotron frequency, $2\pi f_{c} = \frac{eB}{m}$, and  
the Larmor frequency, $2\pi f_{{s}} = \frac{g}{2}\frac{eB}{m}$, respectively.%
\footnote{We set $2\pi f_{{s}} = eB/m$ in the main body.}
Therefore, $g/2$ can be observed by measuring the 
above frequencies or the anomaly frequency defined by
$f_{a} = f_{s} - f_{c}$: 
\begin{equation}
  \frac{g}{2} = \frac{f_{s}}{f_{c}} = 1 + \frac{f_{a}}{f_{c}} \ . \label{gfac}
\end{equation}

\begin{figure}[h]
\centering
\includegraphics[width=10.0cm]{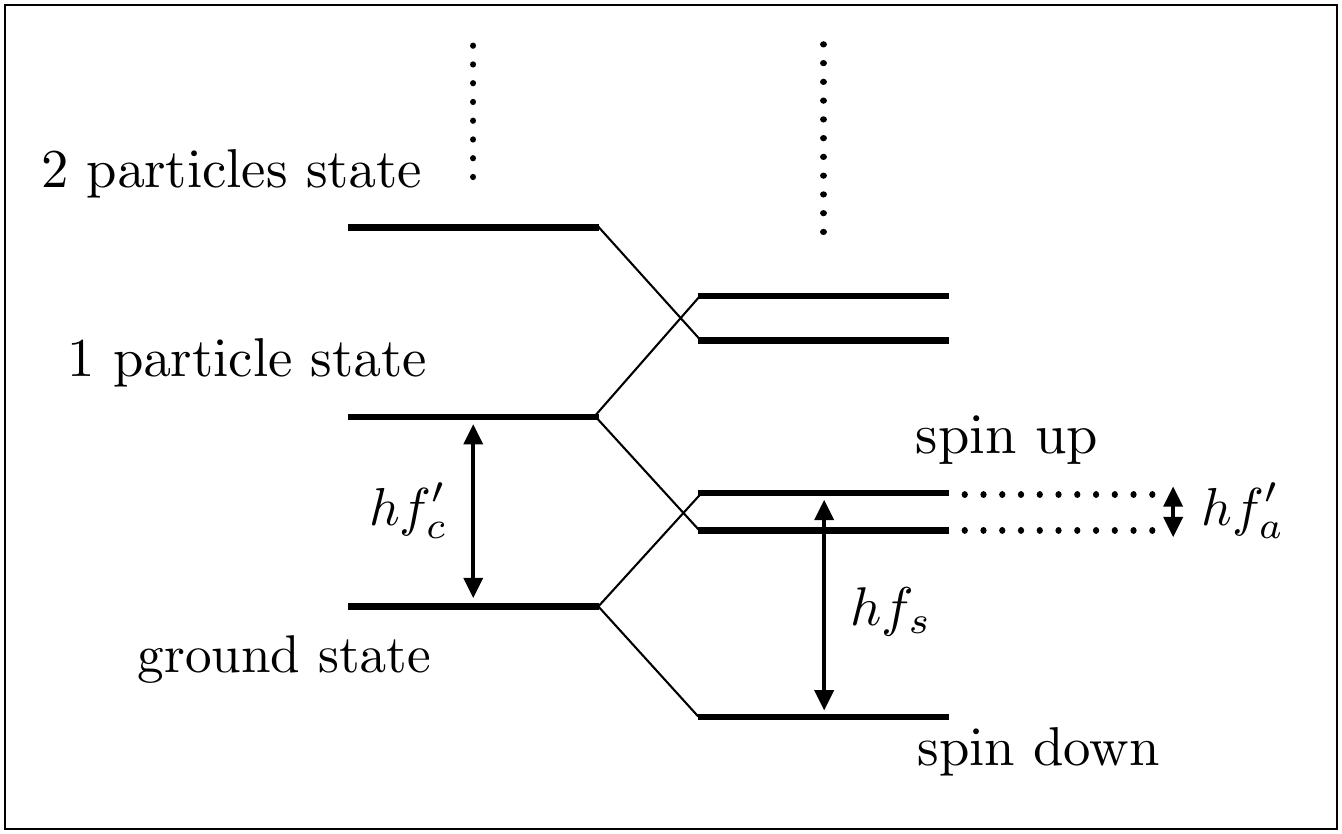}
\caption{The relation of the energy states associated with $f_{c}'$, $f_{s}$ and $f_{a}'$ is illustrated.
The left side ladders represent the energy states of the cyclotron motion.
They are split into the spin up and the spin down states.}
\label{states}
\end{figure}

In turn, let us consider a more realistic situation for measurements.
In the experiment~\cite{Hanneke:2008tm}, an electron is confined in a Penning trap cavity where 
an electrostatic quadrupole potential $V = V_{0} \frac{z^{2} - (x^{2} + y^{2}) }{ 2d^{2} }$ is present
in addition to an external magnetic field $B = (0,0,B)$.
Then, the electron oscillates along with the $z$-direction at an axial frequency, 
$2\pi f_{z} = \sqrt{\frac{eV_{0}}{md^{2}}}$.
The projected motion into $x$-$y$ plane traces an epicyclic orbit, which consists 
of a slow rotation with a large radius at a magnetron frequency, 
$ f_{m} = \frac{ f_{c} - \sqrt{f_{c}^{2} - 2f_{z}^{2} } }{2} $,
and a fast rotation with a small radius at a modified cyclotron frequency,
\begin{equation}
  f_{c}' =  \frac{ f_{c} + \sqrt{f_{c}^{2} - 2f_{z}^{2} } }{2} = f_{c} - f_{m} \ . \label{modc}
\end{equation}
Thus, using a modified anomaly frequency $f_{a}' = f_{s} - f_{c}'$, 
Eq.\,(\ref{gfac}) is rewritten as 
\begin{equation}
  \frac{g}{2} = 1 + \frac{f_{a}' - f_{z}^{2} / 2 f_{c}' } { f_{c}' + f_{z}^{2} / 2 f_{c}'} \ , \label{gg}
\end{equation}
where we have used $f_{m} = f_{z}^{2} / 2 f_{c}'$.
Note that we have omitted a shift of $f_{c}'$ due to special relativistic corrections and 
coupling with cavity modes~\cite{Brown:1985rh,Hanneke:2008tm,Hanneke:2010au} 
because they are not relevant for our purpose. 
One can see that the $g$-factor is determined by measuring $f_{c}'$, $f_{a}'$ and $f_{z}$.
Indeed, in~\cite{Hanneke:2008tm}, above frequencies were measured
directly or indirectly.
The relation among $f_{c}'$, $f_{s}$ and $f_{a}'$ is illustrated in Fig.\,\ref{states}.
We can determine the frequencies by measuring radiated photons from 
energy transitions among the energy levels.
However, typically, the modified cyclotron frequency is $f_{c}' \sim 100$ GHz, 
the modified anomaly frequency is $f_{a}' \sim 100$ MHz and
the axial frequency is $f_{z} \sim 100$ MHz.
Determining energies of radiated photons with high accuracy
around $\sim 100$ GHz is difficult.
Then we are led to operation of quantum jump spectroscopy by monitoring $f_{z}$, which
is measured continuously by detecting current induced by the axial motion.

In order to operate the quantum jump spectroscopy,
we additionally apply a weak magnetic bottle field $B \propto z^{2}$ in the cavity.
Then, the axial motion interacts with the cyclotron motion and the spin precession through the bottle field.
It results in a shift of the axial frequency $f_{z}$ depending on 
the quantum numbers of the cyclotron and the spin energy levels.
Importantly, the interaction Hamiltonian commutes with the cyclotron and the spin Hamiltonian, so that a quantum nondemolition measurement of the cyclotron and the spin states is allowed through monitoring $f_{z}$.
Actually in the measurement~\cite{Hanneke:2008tm}, deriving fields corresponding to 
$f_{c}'$ and $f_{a}'$ are applied in the cavity and then
$f_{c}'$ and $f_{a}'$ can be determined by monitoring time variation of $f_{z}$
due to the cyclotron and the anomaly energy transition.
This quantum jump spectroscopy enables us to determine the $g$-factor 
with very high accuracy.
Remarkably, the current accuracy reaches $2.8 \times 10^{-13}$~\cite{Hanneke:2008tm}.

\bibliography{cyclotron}

\end{document}